\documentclass{pspum-l}

\usepackage[curve]{xypic}

\providecommand{\ch}{\textnormal{ch}}
\providecommand{\Tr}{\textnormal{Tr}}

\providecommand{\Hom}{\textnormal{Hom}}
\providecommand{\Hol}{\textnormal{Hol}}
\providecommand{\cpt}{\textnormal{cpt}}
\providecommand{\Z}{\mathbb{Z}}
\providecommand{\R}{\mathbb{R}}
\providecommand{\odd}{\textnormal{odd}}

\providecommand{\IIm}{\textnormal{Im}}
\providecommand{\CS}{\textnormal{CS}}
\providecommand{\PD}{\textnormal{PD}}

\theoremstyle{definition}

\theoremstyle{remark}

\numberwithin{equation}{section}

\begin{document}

\title{Differential K-characters and D-branes}

\author{Fabio Ferrari Ruffino}
\address{Departamento de Matem\'atica - Universidade Federal de S\~ao Carlos - Rod.\ Washington Lu\'is, Km 235 - C.P.\ 676 - 13565-905 S\~ao Carlos, SP, Brasil}
\email{ferrariruffino@gmail.com}
\thanks{The author was supported by FAPESP, processo 2014/03721-3}




\begin{abstract}
Starting from the definition of Cheeger-Simons $K$-character given in \cite{BM} and \cite{FR}, we show how to describe D-brane world-volumes, the Wess-Zumino action and topological D-brane charges within the $K$-theoretical framework in type II superstring theory. We stress in particular how each feature of the old cohomological classification can be reproduced using the $K$-theoretical language.
\end{abstract}

\maketitle

\section{Introduction}

In the framework of type II superstring theory, there are two fundamental pictures that describe and classify D-brane charges and the Ramond-Ramond fields. The first one relies on classical cohomology. In particular, a D-brane world-volume is a submanifold, which becomes a singular cycle via a suitable triangulation, and the Poincar\'e dual of the underlying homology class is the topological charge. The Ramond-Ramond fields are classified by ordinary differential cohomology, for which the Deligne cohomology provides a concrete model \cite{Brylinski}. The Wess-Zumino action turns out to be the holonomy of a differential cohomology class along the world-volume. The other fundamental classification scheme relies on $K$-theory \cite{Evslin, OS}. In particular, the Ramond-Ramond fields are classified by a differential $K$-theory class \cite{Valentino, MW}, while the topological charge of the D-brane is the corresponding $K$-theory class. What we try to clarify in this paper is how to correctly define the world-volume in this picture, in order to get a suitable generalization of the holonomy map to differential $K$-theory. In this way we are able to give a correct definition of the Wess-Zumino action. Considering the world-volume as a topological $K$-cycle is not enough, thus we have to define a suitable differential extension of $K$-cycles, on which we are able to compute the holonomy. We see that such a definition leads to differential $K$-characters, as defined in \cite{BM} and \cite{FR}. In this way we can draw a complete parallel between the two classification schemes. Since we consider ordinary $K$-theory, we suppose that the $B$-field is vanishing; otherwise, we must develop an analogous construction for twisted $K$-theory and its differential extension.

The paper is organized as follows. In section \ref{OrdinaryDC} we describe the classification scheme via ordinary homology. In section \ref{DiffKRR} we describe the classification scheme via $K$-theory. In section \ref{DiffKH} we recall the definition of differential $K$-character given in \cite{FR}. In section \ref{KHDRR} we apply such a definition in order to describe the world-volume and the Wess-Zumino action in the $K$-theoretical framework, drawing a complete parallel between the two classification schemes.

\section{Ordinary differential cohomology and Ramond-Ramond fields}\label{OrdinaryDC}

If we consider the classical magnetic monopole in 3+1 space-time dimensions, it is well-known that, because of the Dirac quantization condition, the field strength $F_{\mu\nu}$ can be considered as the curvature of a connection on a gauge bundle on $\R^{3} \setminus \{0\}$ (or $\R^{4} \setminus (\{0\} \times \R)$), whose first Chern class, belonging to $H^{2}(\R^{3} \setminus \{0\}; \Z) \simeq \Z$, corresponds to the magnetic charge fixed in the origin. If we argue in the same way for a monopole in a generic space-time dimension $n+1$, we need a gauge invariant integral $(n-1)$-form $F_{\mu_{1}\ldots\mu_{n-1}}$, whose integral on an $(n-1)$-dimensional sphere around the origin of $\R^{n}$ is the magnetic charge (up to a normalization constant). Hence, because of the Dirac quantization condition, such a field strength can be thought of as the curvature of a connection on an abelian $(n-3)$-gerbe, whose first Chern class, belonging to $H^{n-1}(\R^{n} \setminus \{0\}, \Z) \simeq \Z$, corresponds to the charge fixed in the origin. That's why $p$-gerbes naturally arise when dealing with monopoles in a space-time of generic dimension. Since a D-brane, at a semiclassical level, can be thought of as a generalized magnetic monopole whose charge is measured by the Ramond-Ramond field strength, it follows that the Ramond-Ramond potentials $C_{\mu_{1}\ldots\mu_{p+1}}$ and field strength $G_{\mu_{1}\ldots\mu_{p+2}}$ can be thought of respectively as a connection and its curvature on an abelian $p$-gerbe. A concrete way to describe abelian $p$-gerbes with connection is provided by the Deligne cohomology \cite{Brylinski}.

Given a compact smooth manifold $X$, we consider the complex of sheaves:
\begin{equation}\label{ComplexSp}
	S^{p}_{X} = \underline{U}(1) \overset{\tilde{d}}\longrightarrow \Omega^{1}_{\R} \overset{d}\longrightarrow \cdots \overset{d}\longrightarrow \Omega^{p}_{\R},
\end{equation}
where $\underline{U}(1)$ is the sheaf of smooth $U(1)$-valued functions, $\Omega^{k}_{\R}$ is the sheaf of real $k$-forms, $d$ is the exterior differential and $\tilde{d} = \frac{1}{2\pi i} d \circ \log$. The Deligne cohomology group of degree $p$ on $X$ is the $\rm\check{C}$ech hypercohomology group of the complex \eqref{ComplexSp}, i.e., $\check{H}^{p}(X, S^{p}_{X})$. It can be concretely described via a good cover $\mathfrak{U} = \{U_{\alpha}\}_{\alpha \in I}$ of $X$: by definition, we consider the double complex whose columns are the $\rm\check{C}$ech complexes of the sheaves involved in \eqref{ComplexSp}, and we consider the cohomology of the associated total complex. This means that a $p$-cocycle is defined by a sequence $(g_{\alpha_{0} \cdots \alpha_{p+1}}, (C_{1})_{\alpha_{0} \cdots \alpha_{p}}, \ldots, (C_{p})_{\alpha_{0}\alpha_{1}}, (C_{p+1})_{\alpha_{0}})$, satisfying the conditions:
\begin{equation}\label{PGerbesCocycle}
\begin{array}{l}
	(C_{p+1})_{\beta} - (C_{p+1})_{\alpha} = (-1)^{p+1} d(C_{p})_{\alpha\beta}\\
	(C_{p})_{\alpha\beta} + (C_{p})_{\beta\gamma} + (C_{p})_{\gamma\alpha} = (-1)^{p} \, d(C_{p-1})_{\alpha\beta\gamma}\\
	\ldots\\
	\check{\delta}^{p}(C_{1})_{\alpha_{0}\ldots \alpha_{p}} = \frac{1}{2\pi i} d\log g_{\alpha_{0}\ldots \alpha_{p+1}}\\
	\check{\delta}^{p+1}g_{\alpha_{0}\ldots \alpha_{p+1}} = 1.
\end{array}
\end{equation}
The local forms $dC_{p+1}$ glue to a gauge-invariant one $G_{p+2}$, which is the curvature of the $p$-gerbe. We stress that, with respect to this model, the datum of the superstring background must include a complete equivalence class, not only the top-forms $C_{p+1}$. As for line bundles, the correspondence $[G_{p+2}]_{dR} \simeq c_{1}(\mathcal{G}) \otimes_{\Z} \R$ holds, in particular the Dirac quantization condition applies for any $p$. From a physical point of view, Deligne cohomology describes gauge transformations. Conditions \eqref{PGerbesCocycle} specify how the local potentials glue on the intersections, and this concerns a single representative of the equivalence class. There are also possible gauge transformations consisting in the addition of a coboundary. The real datum is the cohomology class, since it is determined by the two real physical observables: the field strength (corresponding to the field $F$ in electromagnetism) and the holonomy of the connection or Wess-Zumino action (corresponding in electromagnetism to the phase difference measured in the context of the Aranhov-Bohm effect). The holonomy is the exponential of the Wilson loop; it can be defined for any $p$ generalizing the definition of the Wilson loop for line bundles. A line bundle with connection is described by a Deligne cohomology class of degree $1$, i.e., by $[(g_{\alpha\beta}, A_{\alpha})] \in \check{H}^{1}(X, S^{1}_{X})$. The Wilson loop is usually described as the minimal coupling between the potentials $A$ and the loop $\gamma$, that's why it is usually written as $\int_{\gamma} A$. Actually the correct definition must also take into account the transition functions. In particular, we divide the loop $\gamma$ in intervals $\gamma_{1}, \ldots, \gamma_{m}$, such that $\gamma_{i}$ is contained in a chart $U_{\alpha_{i}}$. Then we integrate the local potential $A_{\alpha_{i}}$ on $\gamma_{i}$ and we compute the logarithm of the transition function $g_{\alpha_{i}\alpha_{i+1}}$ on the intersection point between $\gamma_{i}$ and $\gamma_{i+1}$. The sum is the Wilson loop, its exponential the holonomy along $\gamma$. Such a definition can be generalized to any $p$, even if the explicit formula is much more complicated to write down concretely \cite{GT}. The basic idea is the following: given a Deligne cohomology class $[(g_{\alpha_{0} \cdots \alpha_{p+1}}, (C_{1})_{\alpha_{0} \cdots \alpha_{p}}, \ldots, (C_{p})_{\alpha_{0}\alpha_{1}}, (C_{p+1})_{\alpha_{0}})]$ of degree $p+1$ and a smooth $(p+1)$-submanifold $\Gamma$, we choose a suitable triangulation of $\Gamma$, such that each simplex is contained in a chart. Then we integrate the potentials $C_{p+1}$ on the $(p+1)$-simplicies, the potentials $C_{p}$ on the $p$-simplicies, and so on until the transition functions on the vertices. A suitable formula joining these data gives the Wilson loop, which is the Wess-Zumino action in string theory. The result depends on the cycle, not only on the homology class, except when the curvature vanishes. This is coherent with the fact that the world-volume is a cycle, not only a homology class. Only in the flat case is the holonomy a morphism from $H_{p+1}(X; \Z)$ to $U(1)$, hence flat abelian $p$-gerbes are classified by the group $H^{p+1}(X; \R/\Z)$. This is due to a Stokes-type formula for the holonomy on a trivial cycle: the holonomy over a boundary $\partial A$ is the exponential of the integral of the curvature on $A$.

Calling $\hat{H}^{p}(X)$ the Deligne cohomology group of degree $p-1$, i.e. $\hat{H}^{p}(X) := \check{H}^{p-1}(X, S^{p-1}_{X})$, we get the following commutative diagram \cite{HS}:
\begin{equation}\label{DiagramDC}
\xymatrix{
	\hat{H}^{\bullet}(X) \ar@{->>}[r]^{c_{1}} \ar@{->>}[d]_{curv} & H^{\bullet}(X; \Z) \ar[d]^{\otimes_{\Z} \R} \\
	\Omega_{int}^{\bullet}(X) \ar[r]^{dR} & H^{\bullet}_{dR}(X).
}
\end{equation}
Here $c_{1}$ is the first Chern class, $curv$ is the curvature, $dR$ is the de-Rham cohomology class and $\Omega_{int}^{\bullet}(X)$ is the group of closed real forms that represent an integral cohomology class. Diagram \eqref{DiagramDC} shows that $\hat{H}^{\bullet}(X)$ is a differential refinement of $H^{\bullet}(X; \Z)$, adding the piece of information due to the connection. Moreover, one can prove that, given a class $\alpha \in \hat{H}^{p+2}(X)$, if $c_{1}(\alpha) = 0$, then $\alpha$ can be represented by a cocycle of the form $(1, 0, \ldots, 0, C_{p+1})$, where $C_{p+1}$ is a globally defined $(p+1)$-form. In this case the Wilson loop on a $(p+1)$-submanifold $\Gamma$ is simply given by $\int_{\Gamma} C_{p+1}$. Such a global potential is unique up to large gauge transformation, i.e., up to the addition of a closed integral form.

With respect to this model, the local Ramond-Ramond potentials $C_{p+1}$ are (a part of) a connection on an abelian $p$-gerbe, whose curvature is the field strength $G_{p+2}$. In this case a D$p$-brane world-volume is thought of as a $(p+1)$-dimensional submanifold $W$ of the space-time $X$. The world-volume $W$, via a suitable triangulation, defines a singular $(p+1)$-cycle, that we also call $W$. When the numerical charge is $q \in \mathbb{Z}$, we think of a stack of $q$ D-branes (anti-branes if $q < 0$), whose underlying cycle is $qW$. The topological charge of the D-brane is the Poincar\'e dual of the underlying homology class $[qW] \in H_{p+1}(X; \mathbb{Z})$. The Wess-Zumino action, usually written as $\int_{W} C_{p+1}$, is the holonomy of the connection on $W$. Moreover, calling $n := \dim\,W$, the violated Bianchi identity is:
	\[dG_{n-p-2} = q \cdot \delta(W) \qquad dG_{p+2} = 0.
\]
This implies that $G_{n-p-2}$ is a closed form in the complement of $W$ and, if $L$ is a linking manifold of $W$, with linking number $l$, we get $\frac{1}{l}\int_{L} G_{n-p-2} = q \in \Z$. That's why field strengths are quantized and can be thought of as the curvature of a connection.

\section{Differential $K$-theory and Ramond-Ramond fields}\label{DiffKRR}

It is well known that $K$-theory is a better tool than ordinary cohomology in order to classify D-brane charges \cite{Evslin, FS}. We first introduce some technical tools about $K$-theory and $K$-homology, then we recall the advantages of the $K$-theoretical classification.

\subsection{$K$-homology} We consider a variant of the usual definition of topological $K$-homology, that will be more suitable for our purposes later: we replace the ``vector bundle modification'' with the Gysin map, which is the natural push-forward in cohomology. We briefly recall the definition. Given an embedding $\iota: Y \rightarrow X$ of compact manifolds of codimension $r$, we consider the following data:
\begin{itemize}
	\item a $K$-orientation of the normal bundle $N_{Y}X$, i.e., a Thom class $u \in K^{r}_{\cpt}(N_{Y}X)$;
	\item a tubular neighbourhood $U$ of $Y$ and a diffeomorphism $\varphi_{U}: N_{Y}X \rightarrow U$;
	\item the open embedding $i: U \hookrightarrow X$, inducing a push-forward in compactly-supported cohomology. Such a push-forward is defined as the pull-back via the map $i': X \rightarrow U^{+}$, which is the identity on $U$ and sends $X \setminus U$ to the point at infinity.
\end{itemize}
There is a natural $K(X)$-module structure on $K_{\cpt}(N_{Y}X)$, hence we define $\iota_{!}: K^{\bullet}(X) \rightarrow K^{\bullet+r}(Y)$ as follows: $\iota_{!}(\alpha) := i_{*}(\varphi_{U})_{*}(\alpha \cdot u)$. The Gysin map turns out to be independent of the choices involved in the construction, except for the orientation of the normal bundle. If $X$ and $Y$ are $K$-oriented manifolds and $\iota$ respects the orientations, since $TX\vert_{Y} \simeq TY \oplus N_{Y}X$, we get an induced orientation on $N_{Y}X$. This implies that the Gysin map is well-defined for an embedding of $K$-oriented manifolds. If $f: Y \rightarrow X$ is a generic smooth map between compact manifolds, we consider an embedding $\iota: Y \hookrightarrow X \times \R^{N}$ such that $\pi_{X} \circ \iota = f$. Then we define $f_{!}(\alpha) := \int_{\R^{N}} \iota_{!}(\alpha)$. Again, if $f$ is a map of $K$-oriented manifolds, we get an induced orientation on $N_{Y}(X \times \R^{N})$, hence the Gysin map is well-defined.

We now come back to $K$-homology. On a smooth compact manifold $X$, we define the group of \emph{$n$-precycles} as the free abelian group generated by the quadruples $(M, u, \alpha, f)$ such that:
\begin{itemize}
	\item $M$ is a smooth compact manifold (without boundary) with $K$-orientation $u$ (i.e., with Thom class $u$ on the tangent bundle), whose connected components $\{M_{i}\}$ have dimension $n+q_{i}$, being $q_{i}$ arbitrary;
	\item $\alpha \in K^{\bullet}(M)$, such that $\alpha\vert_{M_{i}} \in K^{q_{i}}(M)$;
	\item $f: M \rightarrow X$ is a smooth map.
\end{itemize}
We define the group of \emph{$n$-cycles}, denoted by $z_{n}(X)$, as the quotient of the group of $n$-precycles by the free subgroup generated by elements of the form:
\begin{itemize}
	\item $(M, u, \alpha + \beta, f) - (M, u, \alpha, f) - (M, u, \beta, f)$;
	\item $(M, u, \alpha, f) - (M_{1}, u\vert_{M_{1}}, \alpha\vert_{M_{1}}, f\vert_{M_{1}}) - (M_{2}, u\vert_{M_{2}}, \alpha\vert_{M_{2}}, f\vert_{M_{2}})$, for $M = M_{1} \sqcup M_{2}$;
	\item $(M, u, \varphi_{!}\alpha, f) - (N, v, \alpha, f \circ \varphi)$ for $\varphi: (N, v) \rightarrow (M, u)$ a smooth map.
\end{itemize}
We define the group of \emph{$n$-boundaries}, denoted by $b_{n}(X)$, as the subgroup of $z_{n}(X)$ generated by the cycles which are representable by a pre-cycle $(M, u, \alpha, f)$ with the following property: there exists a quadruple $(W, U, A, F)$ such that $W$ is a manifold and $M = \partial W$, $U$ is a $K$-orientation of $W$ and $U\vert_{M} = u$, $A \in K^{\bullet}(W)$ and $A\vert_{M} = \alpha$, $F: W \rightarrow X$ is a smooth map satisfying $F\vert_{M} = f$. We define $K_{n}(X) := z_{n}(X) / b_{n}(X)$. It seems to be more natural to use the Gysin map in the definition, since it is the natural push-forward in cohomology, the vector bundle modification being just a particular case. Moreover, we do not have to quotient out explicitly up to diffeomorphism the first component of the quadruple $(M, u, \alpha, f)$, since the pull-back via a diffeomorphism is again a particular case of the Gysin map. \\

Let us consider a D$p$-brane world-volume $W$ in the space-time $X$. As before we call $n = \dim\,X$. The $U(q)$-gauge theory on $W$ lives on a complex vector bundle $E \rightarrow W$ of rank $q$, being $q$ the number of D-branes in the stack. Hence there is a well-defined $K$-theory class $[E] \in K^{0}(W)$. Moreover, because of the Freed-Witten anomaly \cite{FW}, $W$ is a spin$^{c}$-manifold, which is the condition in order to admit a $K$-theoretical orientation $u$ (that we fix as a part of the world-volume datum). Finally, we consider the embedding in the space-time $\iota: W \hookrightarrow X$. In this way we get a $K$-homology class $[(W, u, E, \iota)] \in K_{p+1}(X)$. Since also $X$ is $K$-orientable (because it is a spin manifold, hence, in particular, spin$^{c}$), we can apply Poincar\'e duality and describe the topological charge as a $K$-theory class of $X$, which is precisely $\iota_{!}[E] \in K^{n-p-1}(X)$.

We can now recall some advantages of the $K$-theoretical classification. First of all, it rules out Freed-Witten anomalous world-volumes, which are precisely the non-$K$-orientable ones. On the contrary, the classification via singular cohomology is unable to detect this anomaly. Moreover, in the $K$-theoretical charge we also take into account the presence of the Chan-Patton bundle and of the embedding in the space-time; this fact will lead to the presence of the gauge and gravitational couplings in the Wess-Zumino action, therefore we get more complete information. Finally, since the D-brane charge is a K-theory class of the space-time, it can be thought of as the formal difference between two space-filling D-brane stacks of equal rank: this is compatible with the Sen conjecture, stating that any D-brane configuration in the space-time can be obtained from a pair made by a D9-brane and a D9-antibrane, via the process of annihilation due to tachyon condensation.

\subsection{Ramond-Ramond fields}

Since the D-brane charge is described by $K$-theory, the Ramond-Ramond fields, that measure such a charge, must be quantized with respect to $K$-theory, not with respect to ordinary cohomology. In order to make this concept more precise, we consider a diagram analogous to \eqref{DiagramDC}, but with respect to $K$-theory instead of ordinary cohomology \cite{BS}. This means that we look for a graded group $\hat{K}^{\bullet}(X)$ fitting into the following diagram: 
\begin{equation}\label{DiagramDK}
\xymatrix{
	\hat{K}^{\bullet}(X) \ar@{->>}[r]^{c_{1}} \ar@{->>}[d]_{curv} & K^{\bullet}(X) \ar[d]^{\ch} \\
	\Omega_{Kint}^{\bullet}(X; \mathfrak{k}_{\R}^{\bullet}) \ar[r]^{dR} & H^{\bullet}_{dR}(X; \mathfrak{k}_{\R}^{\bullet}).
}
\end{equation}
Here $\ch$ is the Chern character, $\mathfrak{k}_{\R}^{\bullet}$ is the $K$-theory ring of the point and $\Omega_{Kint}^{\bullet}(X; \mathfrak{k}_{\R}^{\bullet})$ is the graded group of closed (poly)forms with integral $K$-periods, in the sense that we now specify. Given a form $\omega \in \Omega^{n}_{cl}(X; \mathfrak{k}_{\R}^{\bullet})$ and a $K$-homology class $[(M, u, \alpha, f)] \in K_{n}(X)$, we can consider the following pairing:
\begin{equation}\label{Pairing}
	\langle \omega, [(M, u, \alpha, f)] \rangle := \int_{M} f^{*}\omega \wedge \ch(\alpha) \wedge \hat{A}_{K}(M).
\end{equation}
We say that $\omega$ is $K$-integral or has integral $K$-periods if such a pairing gives an integral value for any $K$-homology class. One can prove that a form is $K$-integral if and only if its cohomology class belongs to the image of the Chern character.

There are various models for $\hat{K}^{\bullet}(X)$; here we consider the Freed-Lott model of $\hat{K}^{0}(X)$ \cite{FL}, that can be extended to any degree (actually only the parity of the degree is meaningful, since Bott periodicity holds even for the differential extension). Given two connections $\nabla$ and $\nabla'$ on the same vector bundle $E$, there is a natural equivalence class $\CS(\nabla, \nabla')$ of odd-dimensional forms up to exact ones, called \emph{Chern-Simons class}, such that $\ch(\nabla) - \ch(\nabla') = d\CS(\nabla, \nabla')$. We define a \emph{differential vector bundle} on $X$ as a quadruple $(E, h, \nabla, \omega)$ where:
\begin{itemize}
	\item $E$ is a complex vector bundle on $X$;
	\item $h$ is an Hermitian metric on $E$;
	\item $\nabla$ is a connection on $E$ compatible with $h$;
	\item $\omega \in \Omega^{\odd}(X)/\IIm(d)$ is a class of real odd-dimensional differential forms up to exact ones.
\end{itemize}
The \emph{direct sum} between differential vector bundles is defined as $(E, h, \nabla, \omega) \oplus (E', h', \nabla', \omega') := (E \oplus E', h \oplus h', \nabla \oplus \nabla', \omega + \omega')$. An \emph{isomorphism of differential vector bundles} $\Phi: (E, h, \nabla, \omega) \rightarrow (E', h', \nabla', \omega')$ is an isomorphism of complex Hermitian vector bundles $\Phi: (E, h) \rightarrow (E', h')$ such that:
\begin{equation}\label{IsomorphismQuadruples}
	\omega - \omega' \in \CS(\nabla, \Phi^{*}\nabla').
\end{equation}
The isomorphism classes of differential vector bundles form an abelian semigroup, hence we can consider its Grothendieck group $\hat{K}^{0}(X)$. By definition an element of $\hat{K}^{0}(X)$ is a difference $[(E, h, \nabla, \omega)] - [(E', h', \nabla', \omega')]$, where $[(E, h, \nabla, \omega)]$ is the class up to the stable equivalence relation.

The group that we have defined fits into the diagram \eqref{DiagramDK} considering the two maps:
	\[c_{1}[(E, h, \nabla, \omega)] := [E] \qquad curv[(E, h, \nabla, \omega)] := \ch(\nabla) - d\omega,
\]
where $\ch(\nabla) = \Tr\exp(\frac{i}{2\pi}\Omega)$, $\Omega$ being the curvature of $\nabla$. The curvature is $K$-integral since:
	\[\langle \ch(\nabla) - d\omega, [(M, u, \alpha, f)] \rangle = \int_{M} \ch(f^{*}E \cdot \alpha) \wedge \hat{A}_{K}(M) \in \mathbb{Z}
\]
because of the index theorem. As we have already pointed out, there is an analogous model for $\hat{K}^{1}(X)$.

We have seen that an abelian $p$-gerbe with vanishing first Chern class can always be represented by a cocycle of the form $(1, 0, \ldots, 0, C_{p+1})$, where $C_{p+1}$ is globally defined and unique up to the addition of an integral form. An analogous consideration holds for differential $K$-theory: a class $\alpha \in \hat{K}^{p}(X)$ with vanishing first Chern class can be represented in the form $[(0, 0, 0, C_{p-1})]$, where $C_{p-1} \in \Omega^{p-1}(X; \mathfrak{k}_{\R}^{\bullet})$ is globally defined and unique up to the addition of a $K$-integral form. \\

We are now able to describe the Ramond-Ramond fields in the $K$-theoretical framework. Because of the Bott periodicity, the two meaningful groups are $\hat{K}^{0}(X)$ and $\hat{K}^{1}(X)$, corresponding respectively to type IIB and type IIA theory. The Ramond-Ramond fields with even-degree field strength are jointly classified by a class $\alpha \in \hat{K}^{0}(X)$, while the ones with odd-degree field strength are classified by $\beta \in \hat{K}^{1}(X)$. We discuss the features of $\alpha$, the discussion about $\beta$ being analogous. The curvature of $\alpha$ is a form $G_{ev} \in \Omega^{0}_{cl}(X; \mathfrak{k}^{\bullet}_{\mathbb{R}}) \simeq \bigoplus_{p \in \mathbb{Z}} \Omega^{2p}_{cl}(X)$. The component of degree $2p$ is the field-strength $G_{2p}$. If we consider a local chart $U$ of $X$, then $\alpha\vert_{U}$ is topologically trivial, hence it can be represented in the form $(0, 0, 0, C_{odd})$, with $C_{odd} \in \Omega^{-1}(U; \mathfrak{k}^{\bullet}_{\mathbb{R}}) \simeq \bigoplus_{p \in \mathbb{Z}} \Omega^{2p-1}(U)$, unique up to the addition of an exact form (on a contractible chart $U$, any $K$-integral form, being closed, is exact). The component of degree $2p-1$ is the local potential $C_{2p-1}$. This means that the potentials are a local expression of a global differential $K$-theory class, which is the complete datum encoded in the space-time.

Now the main point is the following. How do we have to think of a D-brane world-volume in the $K$-theoretical framework, in order to correctly define the Wess-Zumino action? Comparing with the framework of ordinary cohomology, it seems natural to think of it as a $K$-homology cycle, representing a class whose Poincar\'e dual is the topological charge. This is possible, but we will see that it is not enough in order to define the Wess-Zumino action.

\subsection{Comparing the two frameworks}

Let us start from the mathematics. In table \ref{fig:Comparison} we compare the features of ordinary differential cohomology with the ones of differential $K$-theory.
\begin{table*}
	\centering
		\begin{tabular}{|l|l|l|}
			\hline & & \\ & \textbf{Abelian $p$-gerbe with c.} & \textbf{Diff.\ $K$-theory class} \\ & & \\ \hline
			& & \\ \textbf{Classified by} & $\hat{H}^{\bullet}(X)$ & $\hat{K}^{\bullet}(X)$ \\ & & \\ \hline
			& & \\ \textbf{First Chern class} & $c_{1} \in H^{\bullet}(X; \Z)$ & $c_{1} \in K^{\bullet}(X)$ \\ & & \\ \hline
			& & \\ \textbf{Curvature} & $curv \in \Omega^{\bullet}_{int}(X)$ & $curv \in \Omega^{\bullet}_{Kint}(X; \mathfrak{k}_{\R}^{\bullet})$ \\  & & \\ & $[curv]_{dR} \simeq c_{1} \otimes_{\Z} \R$ & $[curv]_{dR} \simeq \ch(c_{1})$ \\ & & \\ \hline
			& & \\ \textbf{Top.\ trivial classes} & $\Omega^{\bullet-1}(X)/\Omega^{\bullet-1}_{int}(X)$ & $\Omega^{\bullet-1}(X; \mathfrak{k}_{\R}^{\bullet})/\Omega^{\bullet-1}_{Kint}(X; \mathfrak{k}_{\R}^{\bullet})$ \\ & & \\ \hline
			& & \\ \textbf{Flat classes} & $H^{\bullet-1}(X; \R/\Z)$ & $K^{\bullet-1}(X; \R/\Z)$ \\ & & \\ \hline
			& & \\ \textbf{Holonomy} & $Z_{\bullet-1}^{sm}(X) \rightarrow U(1)$ & $?? \rightarrow U(1)$ \\ & & \\ \hline
		\end{tabular}
\caption{Comparison}\label{fig:Comparison}
\end{table*}
We can see that there is a complete analogy between the two pictures, except for the holonomy, since we have to clarify on which cycles it must be computed in the case of $K$-theory (in the table, $Z^{sm}_{\bullet}$ denotes the smooth singular cycles).

Physically, Ramond-Ramond fields in type II superstring theory are classified by an abelian $p$-gerbe or by a differential $K$-theory class (line 1 of table \ref{fig:Comparison}). The field strength is the curvature in each case, hence it obeys the corresponding quantization condition (line 3 of table \ref{fig:Comparison}). Any class is locally topologically trivial, hence we get the local Ramond-Ramond potentials up to gauge transformations (line 4 of table \ref{fig:Comparison}). The world-volume is a singular cycle in the first picture, and the Poincar\'e dual of the underlying homology class is the topological charge; the Wess-Zumino action is the holonomy of the Ramond-Ramond fields on the world-volume (line 6 of table \ref{fig:Comparison}). What is the Wess-Zumino action in the $K$-theoretical framework?

We have seen that the topological D-brane charge is measured by the $K$-theory class of the space-time Poincar\'e dual to $[(W, u, E, \iota)] \in K_{p+1}(X)$, where $W$ is the world-volume as a sub-manifold, $u$ is a fixed Thom class of the tangent bundle of $W$, $E$ is the Chan-Patton bundle and $\iota$ is the embedding of $W$ in the space-time. This class is $\iota_{!}[E]$. Hence, we could consider as the world-volume the $K$-cycle $(W, u, E, \iota)$, but we do not know how to define the holonomy of the class $\alpha \in \hat{K}^{p+2}(X)$, representing the Ramond-Ramond fields. Usually the pairing is written supposing that $\alpha$ is topologically trivial, hence described by a global form $C$. It has the following form \cite{MM}:
\begin{equation}\label{PairingC}
	\langle \alpha, (W, u, E, \iota) \rangle = \int_{W} C \wedge \ch(E) \wedge \hat{A}_{K}(W) \wedge \hat{A}_{K}(X)^{-\frac{1}{2}}.
\end{equation}
We denote by $\hat{A}_{K}$ the $\hat{A}$-genus of $K$-theory, i.e., $\hat{A} \wedge e^{\frac{d}{2}}$, where $d \in H^{2}(W; \Z)$ is a suitable class whose $\Z_{2}$-reduction is $w_{2}(W)$ \cite{MM}. Equation \eqref{PairingC} has some problems. The most evident one is what we have already said: it holds only when $\alpha$ is topologically trivial. Actually, even in this case, we can make some more remarks. The form $C$ in general is not-closed, hence the integral on $W$ depends on the specific representatives of $\ch(E)$ and $\hat{A}_{K}(W)$ (we neglect for the moment $\hat{A}_{K}(X)$, since it does not depend on the D-brane). How do we choose them? It is not difficult to reply for $\ch(E)$: since $\ch(E) = [\Tr\exp(\frac{i}{2\pi}\Omega)]$, $\Omega$ being the curvature of a connection on $E$, we have to fix a connection on $E$ in order to fix a representative of $\ch(E)$. We choose the connection defining the $U(q)$-gauge theory on the D-brane, $q$ being the rank of $E$. This fact shows that, even when $\alpha$ is topologically trivial, we cannot consider as the world-volume the topological $K$-cycle $(W, u, E, \iota)$: at least we need to include the connection on $E$ as a part of the datum. Moreover, what about the $\hat{A}$-genus? It does not seem so trivial to find a natural representative, hence we need some information more. In the next sections we try to fill this gap.

There is an element missing in the previous list: in the framework of ordinary cohomology, the numerical charge of a D-brane is measured by the integral of the dual field-strength on a linking manifold. It is not difficult to find the analogous property in the $K$-theoretical framework, actually we could do this even considering the world-volume just as a topological $K$-cycle, but we postpone the discussion to the last section.

\section{Differential $K$-characters}\label{DiffKH}

We try to reply to the previous questions looking for a suitable definition of differential $K$-cycle and differential $K$-character. The idea we presented in \cite{FR} is the following. Let us consider a $K$-cycle $(M, u, \alpha, f)$ of degree $p$ on $X$ and a differential $K$-theory class $\hat{\beta} \in \hat{K}^{p+1}(X)$ (of course only the parity of $p$ is meaningful). We have that $\alpha \in K^{q}(X)$, where $q$ satisfies $\dim M = p + q$. If we refine $\alpha$ to a differential class $\hat{\alpha}$, then we can consider the product $\hat{\alpha} \cdot f^{*}\hat{\beta} \in \hat{K}^{p+q+1}(M)$. There is a unique map from $M$ to the point, that we call $p_{M}$. If we are able to define the differential refinement of the Gysin map, via a suitable differential refinement of the orientation $u$ (that we call $\hat{u}$), we can calculate $(p_{M})_{!}(\hat{\alpha} \cdot f^{*}\hat{\beta}) \in \hat{K}^{1}(pt)$. We now prove that $\hat{K}^{1}(pt) \simeq \R/\Z$ canonically, hence we can define the holonomy of $\hat{\beta}$ on $(M, \hat{u}, \hat{\alpha}, f)$ as $\exp((p_{M})_{!}(\hat{\alpha} \cdot f^{*}\hat{\beta}))$. This shows that, in order to define the holonomy, we must consider a suitable differential refinement of the topological $K$-cycles, that will lead us to define differential $K$-characters. We have to show that $\hat{K}^{1}(pt) \simeq \R/\Z$ canonically. Since $K^{1}(pt) = 0$, a class $\gamma \in \hat{K}^{1}(pt)$ is topologically trivial, hence it can be represented by a form $\omega \in \Omega^{0}(pt; \mathfrak{k}^{\bullet}_{\R})/\Omega^{0}_{Kint}(pt; \mathfrak{k}^{\bullet}_{\R})$. On a point there are non-zero forms only in degree $0$, and they are real numbers. The $K$-integral ones are precisely the integer numbers, since, in the pairing \eqref{Pairing}, $f^{*}\omega$ is constant and $\int_{M} \ch(\alpha) \wedge \hat{A}_{K}(M)$ is integral because of the index theorem. This shows that $\Omega^{0}(pt; \mathfrak{k}^{\bullet}_{\R})/\Omega^{0}_{Kint}(pt; \mathfrak{k}^{\bullet}_{\R}) \simeq \R/\Z$. \\

Let us present the precise definition of differential $K$-character. We have shown above that we must consider suitable differential refinements of the components of a topological $K$-cycle. The main point is that, when dealing with differential classes, the curvature is meaningful as a single form, not only as a cohomology class, therefore it is not homotopy invariant. Thus, we need suitable definitions in order to recover classical topological tools as the 2x3 rule about the orientation of the bundles $E$, $F$, $E \oplus F$. In particular, we have to correctly define the concept of orientation of a smooth map with respect to differential $K$-theory \cite{HS}, which encodes the data that we need to fix. First of all, following \cite{Bunke}, we define a \emph{$\hat{K}$-orientation} of a smooth vector bundle as a differential extension\footnote{A differential extension of a class $\alpha \in K^{n}(X)$ is a class $\hat{\alpha} \in \hat{K}^{n}(X)$ such that $c_{1}(\hat{\alpha}) = \alpha$.} of a Thom class of the bundle. Then we define a \emph{representative of a $\hat{K}$-orientation} of a smooth map $f: Y \rightarrow X$ between compact manifolds (neat if $X$ and $Y$ have boundary) as the datum of:
\begin{itemize}
	\item a (neat) embedding $\iota: Y \hookrightarrow X \times \R^{N}$ for any $N \in \mathbb{N}$, such that $\pi_{X} \circ \iota = f$;
	\item a $\hat{K}$-orientation $\hat{u}$ of the normal bundle $N_{Y}(X \times \R^{N})$;
	\item a (neat) tubular neighbourhood $U$ of $Y$ in $X \times \R^{N}$ with a diffeomorphism $\varphi: N_{Y}(X \times \R^{N}) \rightarrow U$.
\end{itemize}
Using a definition similar to the topological one, it turns out that the Gysin map $f_{!}: \hat{K}^{\bullet}(Y) \rightarrow \hat{K}^{\bullet+r}(X)$ is well defined if $f$ is endowed with a representative of a $\hat{K}$-orientation. We can suitably define \emph{homotopy} and \emph{equivalence by stabilization} in the set of representatives of $\hat{K}$-orientations, and we call \emph{$\hat{K}$-orientation} of $f$ an equivalence class. Moreover, a smooth manifold $M$ is \emph{$\hat{K}$-oriented} if the unique map from $M$ to a point is $\hat{K}$-oriented.

With this definition, as in the topological case, if $f$ is a \emph{proper submersion} between $\hat{K}$-oriented manifolds, then it automatically inherits an orientation. Actually the technical details are more complicated. We just sketch the problems. First of all, one fundamental property of the Gysin map is that it is compatible with the composition, i.e., $(g \circ f)_{!} = g_{!} \circ f_{!}$. Moreover, it satisfies $f_{!}(\alpha \cdot f^{*}\beta) = f_{!}\alpha \cdot \beta$. In order to maintain these properties in the differential case, we need the hypothesis that $f$ is a submersion, because, in this case, considering the embedding $\iota: Y \hookrightarrow X \times \R^{N}$, we can choose the tubular neighbourhood of $Y$ in such a way that the image of the fibre of the normal bundle on $y \in Y$ is contained in $\{\iota(y)\} \times \R^{N}$. In this way, when we consider $\alpha \cdot f^{*}\beta$ and we apply $f_{!}$, the multiplication by $\beta$ acts as a multiplication by a constant class on each fibre of the tubular neighbourhood, therefore it factorizes in the integral with respect to $\R^{N}$. A similar argument holds in order to prove that $(g \circ f)_{!} = g_{!} \circ f_{!}$. Moreover, thanks to the equivalence relation we introduced among the representatives of orientations, the embedding $\iota$ is meaningful only up to homotopy and stabilization, and the choice of the tubular neighborhood is immaterial. This is important by a physical point of view, since a fixed embedding and a fixed tubular neighbourhood would have no physical meaning.

Now we can come back to the definition of differential $K$-character. On a smooth compact manifold $X$, we define the group of \emph{differential $n$-precycles} as the free abelian group generated by the quadruples $(M, \hat{u}, \hat{\alpha}, f)$ such that:
\begin{itemize}
	\item $M$ is a smooth compact manifold (without boundary) with $\hat{K}^{\bullet}$-orientation $\hat{u}$,\footnote{Here we denote by $\hat{u}$ the whole differential orientation, not only the differential refinement of the Thom class $u$.} whose connected components $\{M_{i}\}$ have dimension $n+q_{i}$, with $q_{i}$ arbitrary;
	\item $\hat{\alpha} \in \hat{K}^{\bullet}(M)$, such that $\hat{\alpha}\vert_{M_{i}} \in \hat{K}^{q_{i}}(M)$;
	\item $f: M \rightarrow X$ is a smooth map.
\end{itemize}
The group of \emph{differential $n$-cycles}, denoted by $\hat{z}_{n}(X)$, is the quotient of the group of $n$-precycles by the free subgroup generated by elements of the form:
\begin{itemize}
	\item $(M, \hat{u}, \hat{\alpha} + \hat{\beta}, f) - (M, \hat{u}, \hat{\alpha}, f) - (M, \hat{u}, \hat{\beta}, f)$;
	\item $(M, \hat{u}, \hat{\alpha}, f) - (M_{1}, \hat{u}\vert_{M_{1}}, \hat{\alpha}\vert_{M_{1}}, f\vert_{M_{1}}) - (M_{2}, \hat{u}\vert_{M_{2}}, \hat{\alpha}\vert_{M_{2}}, f\vert_{M_{2}})$, for $M = M_{1} \sqcup M_{2}$;
	\item $(M, \hat{u}, \varphi_{!}\hat{\alpha}, f) - (N, \hat{v}, \hat{\alpha}, f \circ \varphi)$ for $\varphi: N \rightarrow M$ a submersion, oriented via the 2x3 principle.
\end{itemize}
The group of \emph{differential $n$-boundaries}, denoted by $\hat{b}_{n}(X)$, is the subgroup of $\hat{z}_{n}(X)$ generated by the cycles which are representable by a pre-cycle $(M, \hat{u}, \hat{\alpha}, f)$ with the following property. There exists a quadruple $(W, \hat{U}, \hat{A}, F)$ such that $W$ is a manifold and $M = \partial W$, $\hat{U}$ is an $\hat{K}^{\bullet}$-orientation of $W$ and $\hat{U}\vert_{M} = \hat{u}$, $\hat{A} \in \hat{K}^{\bullet}(W)$ and $\hat{A}\vert_{M} = \hat{\alpha}$, $F: W \rightarrow X$ is a smooth map satisfying $F\vert_{M} = f$.
We define $K_{n}(X) := \hat{z}_{n}(X) / \hat{b}_{n}(X)$.

The homology groups, defined in this way, are isomorphic to the ones defined via topological cycles, as shown above. We have defined the differential cycles in such a way that it is possible to integrate a differential cohomology class on such a cycle. When the class is flat and only the homology class is meaningful, we need no differential information, since the group of flat classes is $\Hom(K_{p-1}(X), \R/\Z)$; that's why we do not need a non-trivial differential extension of the homology classes. We will see in the following the physical meaning of this fact. \\

Given a class $\hat{\beta} \in \hat{K}^{p+1}(X)$ and a differential $p$-cycle $(M, \hat{u}, \hat{\alpha}, f)$, with $\dim M = p + q$ and $\hat{\alpha} \in \hat{K}^{q}(M)$, we can compute the holonomy as we sketched at the beginning of this paragraph: we consider the class $\hat{\alpha} \cdot f^{*}\hat{\beta} \in \hat{K}^{p+q+1}(M)$ and, since $M$ is $\hat{K}$-oriented, i.e., the map $p_{M}: M \rightarrow pt$ is $\hat{K}$-oriented, we can calculate $(p_{M})_{!}(\hat{\alpha} \cdot f^{*}\hat{\beta}) \in \hat{K}^{1}(pt) \simeq \R/\Z$. The exponential of the result is the holonomy. One can show that the holonomy completely characterizes the differential $K$-theory class, as in the case of ordinary cohomology. When the cycle is a boundary, a Stokes-type formula holds even in the $K$-theoretical framework: if $(M, \hat{u}, \hat{\alpha}, f) = \partial(W, \hat{U}, \hat{A}, F)$, then
\begin{equation}\label{StokesK}
	\Hol_{(M, \hat{u}, \hat{\alpha}, f)}(\hat{\beta}) = \exp\int_{W} F^{*}curv(\hat{\beta}) \wedge curv(\hat{A}) \wedge \hat{A}_{\hat{K}}(W).
\end{equation}
Here $\hat{A}_{\hat{K}}(W)$ is a representative of $\hat{A}_{K}(W)$, which is defined as $\int_{N_{W}\R^{N}/W} curv(\hat{u})$, where the embedding of $W$ in $\R^{N}$ is provided by the differential orientation of $W$. Formula \eqref{StokesK} implies that, if $\alpha$ is flat, its holonomy over a trivial cycle is zero. Hence, in this case, the holonomy only depends on the $K$-homology class.

Thanks to differential $K$-characters we can complete table \ref{fig:Comparison}: in the $K$-theoretical framework, the holonomy is a group morphism $\hat{z}_{\bullet-1}(X) \rightarrow U(1)$.

\section{Differential $K$-characters, D-branes and Ramond-Ramond fields}\label{KHDRR}

We can now complete the $K$-theoretical description of D-branes in type II superstring theory. We describe a D-brane world-volume as a differential $K$-cycle. In particular, we consider the topological $K$-cycle $(W, u, E, \iota)$, where (we recall) $u$ is a Thom class of $W$, $E$ is the Chan-Patton bundle and $\iota$ is the embedding of $W$ in the space-time. On $E$ there is the $U(q)$-gauge theory of the D-brane, hence $E$ is endowed with an Hermitian metric $h$ and a compatible connection $\nabla$. Therefore, we can consider the differential $K$-theory class $[(E, h, \nabla, 0)]$, using the Freed-Lott model. We call $\hat{E}$ such a class. Moreover, we refine $u$ to a differential orientation $\hat{u}$ of $W$, that must be fixed as a part of the datum. We get a differential $K$-theory class $(W, \hat{u}, \hat{E}, \iota)$, that is the world-volume in the $K$-theoretical framework. Actually, we consider one cycle made by all the even-dimensional world-volumes or one made by all the odd-dimensional ones, depending whether we are considering the type IIA or type IIB theory. In this way, we can correctly define the Wess-Zumino action: it is the holonomy of the differential $K$-theory class, representing the Ramond-Ramond fields, on the world-volume. How do we compute the topological charge? Here we see the physical importance of the fact that the $K$-homology groups, defined via differential cycles and boundaries, are isomorphic to the ones defined via topological cycles and boundaries: the Poincar\'e dual of the underlying $K$-homology class of the world-volume is the topological charge that we have already defined.

We show that, when the class is topologically trivial, the holonomy coincides with \eqref{PairingC}. Actually, we obtain this result normalizing the class with $\hat{A}_{\hat{K}}(X)^{-\frac{1}{2}}$. In fact, let us call $a(C_{odd})$ the topologically trivial class represented by the global form $C_{odd}$, i.e., in the Freed-Lott model, $a(C_{odd}) = [(0, 0, 0, C_{odd})]$ (again, the discussion about $C_{ev}$ is analogous). Let us compute the holonomy of $a(C_{odd} \wedge \hat{A}_{\hat{K}}(X)^{-\frac{1}{2}})$ along the differential $K$-cycle $(W, \hat{u}, \hat{E}, \iota)$. We have that:
\begin{equation}\label{HolTopTrivial}
\begin{split}
	(p_{W})_{!}(\iota^{*}a(C_{odd} \wedge &\hat{A}_{\hat{K}}(X)^{-\frac{1}{2}}) \cdot \hat{E}) = (p_{W})_{!}(a(\iota^{*}C_{odd} \wedge \hat{A}_{\hat{K}}(X)^{-\frac{1}{2}} \\
	& \wedge curv(\hat{E})) = (p_{W})_{!}(a(\iota^{*}C_{odd} \wedge \hat{A}_{\hat{K}}(X)^{-\frac{1}{2}} \wedge \ch\nabla_{E})).
\end{split}
\end{equation}
For the first equality we have used the relation $a(C_{odd}) \cdot \hat{E} = a(C_{odd} \wedge curv(\hat{E}))$, which is a fundamental property of differential cohomology. Now we apply the definition of the Gysin map. We consider the data provided by any representative of the differential orientation $\hat{u}$ of $W$: an embedding $j: W \hookrightarrow \R^{N}$, a tubular neighbourhood $U$ of $W$ in $\R^{N}$, the diffeomorphism $\varphi_{U}: N_{W}\R^{N} \rightarrow U$ and the open embedding $i: U \hookrightarrow \R^{N}$. From \eqref{HolTopTrivial} we get:
	\[\begin{split}
	\int_{\R^{N}} i_{*} (\varphi_{U}&)_{*}(a(\iota^{*}C_{odd} \wedge \hat{A}_{\hat{K}}(X)^{-\frac{1}{2}} \wedge \ch\nabla_{E}) \cdot \hat{u}) \\
	& = \int_{\R^{N}} i_{*} (\varphi_{U})_{*}(a(\iota^{*}C_{odd} \wedge \hat{A}_{\hat{K}}(X)^{-\frac{1}{2}} \wedge \ch\nabla_{E} \wedge curv(\hat{u}))) \\
	& = a\biggl(\int_{N_{W}\R^{N}} \iota^{*}C_{odd} \wedge \hat{A}_{\hat{K}}(X)^{-\frac{1}{2}} \wedge \ch\nabla_{E} \wedge curv(\hat{u})\biggr) \\
	& = a\biggl(\int_{W} \int_{N_{W}\R^{N}/W} \iota^{*}C_{odd} \wedge \hat{A}_{\hat{K}}(X)^{-\frac{1}{2}} \wedge \ch\nabla_{E} \wedge curv(\hat{u})\biggr) \\
	& = a\biggl(\int_{W} \iota^{*}C_{odd} \wedge \hat{A}_{\hat{K}}(X)^{-\frac{1}{2}} \wedge \ch\nabla_{E} \wedge \int_{N_{W}\R^{N}/W} curv(\hat{u})\biggr) \\
	& = a\biggl(\int_{W} \iota^{*}C_{odd} \wedge \ch\nabla_{E} \wedge \hat{A}_{\hat{K}}(W) \wedge \hat{A}_{\hat{K}}(X)^{-\frac{1}{2}} \biggr).
\end{split}\]
Thus the holonomy is the exponential of $\int_{W} \iota^{*}C_{odd} \wedge \ch\nabla_{E} \wedge \hat{A}_{\hat{K}}(W) \wedge \hat{A}_{\hat{K}}(X)^{-\frac{1}{2}}$, as stated in equation \eqref{PairingC}. We see that, in this case, we have canonical representatives of $\ch E$ and $\hat{A}_{K}(X)$, provided by the curvatures of $\hat{E}$ and $\hat{u}$, that are two components of the world-volume thought of as a differential $K$-cycle. Since it is necessary to normalize with $\hat{A}_{\hat{K}}(X)^{-\frac{1}{2}}$ the $K$-theory class whose holonomy we are calculating, we have to fix a representative of such a class as a part of the background. This would follow automatically refining the space-time manifold to a differential $K$-cycle too, but it is not necessary, we just choose a representative of the $\hat{A}$-genus as a normalization constant. \\

Using classical cohomology, the integral of the field-strength along a linking manifold is the numerical charge of the D-brane. A linking manifold $L$ of $W$ is the boundary of a manifold $S$ that intersects $W$ transversely in a finite number of points of the interior. The number of such points is the linking number. Within the $K$-theoretical framework, we can generalize this concept. First of all, when we consider a D-brane world-volume $W$ with Chan-Patton bundle $E$, there is not only that charge of $W$ itself, but there are sub-brane charges which are encoded in $E$. In particular, since in the Wess-Zumino action the Chern character $\ch\,E$ appears, we can interpret the Poincar\'e duals of the Chern characters as sub-branes of $W$ with a charge. Because of this, a linking manifold of $W$ is not enough. Since the field-strength is $K$-quantized, it is natural to consider a \emph{linking $K$-cycle} $(L, u, F, \iota)$. Here $L$ is a ``generalized'' linking manifold, i.e., $L$ is the boundary of a manifold $S$ such that $S$ and $W$ intersect transversely in a submanifold (without boundary) contained in the interior of $S$. If $S \cap W$ is $0$-dimensional, we get a linking manifold in the usual sense. We consider the even-dimensional field-strengths $G_{ev}$, the discussion about $G_{odd}$ being analogous. The violated Bianchi identity is \cite{MM}:
\begin{equation}\label{BianchiK}
	dG_{ev} = \delta(W) \wedge \ch\nabla_{E} \wedge \hat{A}_{\hat{K}}(W) \wedge \hat{A}_{\hat{K}}(X)^{-\frac{1}{2}},
\end{equation}
where $W$ is the union of all the world-volumes with dimension of the suitable parity. Here, again, we see the importance of having representatives of the Chern character and the $\hat{A}$-genus, because $dG_{ev}$ is a form (actually, a current) and not a cohomology class. Equation \eqref{BianchiK} implies that $G_{ev} \wedge \hat{A}_{\hat{K}}(X)^{-\frac{1}{2}}$ is $K$-quantized and the pairing with a linking $K$-cycle gives the corresponding charge. In fact:
	\[\begin{split}
	\langle G_{ev} \wedge \hat{A}_{K}(X)^{-\frac{1}{2}}, &(L, u, F, \iota) \rangle = \int_{L} G_{ev} \wedge \hat{A}_{K}(X)^{-\frac{1}{2}} \wedge \ch(F) \wedge \hat{A}_{K}(L) \\
	& = \int_{S} dG_{ev} \wedge \hat{A}_{\hat{K}}(X)^{-\frac{1}{2}} \wedge \ch(F) \wedge \hat{A}_{\hat{K}}(S) \\
	& = \int_{S} \delta(W) \wedge \ch(E \otimes F) \wedge \frac{\hat{A}_{\hat{K}}(W) \wedge \hat{A}_{\hat{K}}(S)}{\hat{A}_{\hat{K}}(X)} \\
	& = \int_{S \cap W} \ch(E \otimes F) \wedge \hat{A}_{K}(S \cap W) \in \mathbb{Z}.
\end{split}\]
If $L$ is a linking manifold and $F$ is the trivial line bundle, then we get $\int_{S \cap W} \ch^{0}E = ql$, as in the previous case ($l$ is the linking number and $q = \ch^{0}E$). Let us consider $\ch^{1}E$. If we represent $PD_{W}(\ch^{1}E)$ as a cycle $qW'$ of codimension $2$, we suppose that we can take a linking manifold of $W'$, such that $S \cap W$ is a submanifold of dimension $2$. Then the corresponding term of the integral is $\int_{S \cap W} \ch^{1}E = \int_{qW'} 1 = ql$, i.e., we measure the charge of the sub-brane. An analogous consideration holds for the higher Chern characters, but we have to take into account the terms of the $\hat{A}$-genus. We just make two final remarks. Using ordinary cohomology, in order to compute the linking number $l$ we must consider any solution of $dG_{n-p-2} = \delta(W)$ (with $q = 1$) and compute the integral along $L$. Similarly, in the $K$-theoretical picture, in order to compute the linking number of a cycle $(L, u, F, \iota)$, we consider any solution of $dG_{ev} = \delta(W) \wedge \hat{A}_{\hat{K}}(W) \wedge \hat{A}_{\hat{K}}(\hat{X})^{-\frac{1}{2}}$ (with $E$ the trivial line bundle) and compute the integral along the cycle. Then, from the previous integral, we can compute $q$. Moreover, we remark that the fact that $G_{ev} \wedge \hat{A}_{\hat{K}}(X)^{-\frac{1}{2}}$, and not $G_{ev}$ itself, is $K$-quantized, is just a normalization analogous to $\frac{1}{2\pi}G_{p}$ in the case of ordinary cohomology (the constant can appear depending on the conventions). Here $\hat{A}_{\hat{K}}(X)$ does not depend on $W$, hence it is a constant with respect to a fixed space-time background. \\

Now we have all the elements in order to draw a complete parallel between the two classification schemes of D-branes. Table \ref{fig:ComparisonPhys} shows such a parallel.

\begin{table*}[h!]
	\centering
		\begin{tabular}{|l|l|l|}
			\hline & & \\ & \textbf{Singular cohomology} & \textbf{$K$-theory} \\ & & \\ \hline
			& & \\ \textbf{World-vol.} & Singular cycle $qW$ & Diff.\ $K$-cycle $(W, \hat{u}, \hat{E}, \iota)$ \\ & & \\ \hline
			& & \\ \textbf{Top.\ charge} & Sing.\ coh.\ class $\PD_{X}[qW]$ & $K$-th.\ class $\PD_{X}[(W, \hat{u}, \hat{E}, \iota)]$ \\ & & \\ \hline
			& & \\ \textbf{RR fields} & Ordinary diff.\ cohom.\ class & Diff.\ $K$-theory class \\  & & \\ & Integral field strength & $K$-Integral field strength \\ & & \\ \hline
			& & \\ \textbf{WZ action} & Holonomy of the RR fields & $K$-Holonomy of the RR fields \\ & & \\ \hline
			& & \\ \textbf{Num.\ charge} & $\int$ f.s.\ over a linking manifold & $\int$ f.s.\ over a linking $K$-cycle \\ & & \\ \hline
		\end{tabular}
\caption{Comparison (physics).}\label{fig:ComparisonPhys}
\end{table*}


\bibliographystyle{amsalpha}

\end{document}